\documentclass[
  ,draft            
  ]
  {aipproc}

\layoutstyle{6x9}

\begin{document}

\title{A Bayesian Analysis of Extrasolar Planet Data for HD 208487}


\author{P. C. Gregory}{
  address={Physics and Astronomy Department, University of British Columbia,\\
    Vancouver, BC V6T 1Z1}
}{\footnotesize \noindent For publication in ``Bayesian Inference and Maximum Entropy Methods'', San Jose 2005,\\ K. H. Knuth, A. E. Abbas, R. D. Morris, J. P. Castle (eds.), AIP Conference Proceeding}

\begin{abstract}
{Precision radial velocity data for HD 208487 has been re-analyzed using a new Bayesian multi-planet Kepler periodogram. The periodgram employs a parallel tempering Markov chain Monte Carlo
algorithm with a novel statistical control system. We confirm the previously reported orbit (Tinney et al. 2005) of $130$ days. In addition, we conclude there is strong evidence for a second planet with a period of $998_{-62}^{+57}$ days, an eccentricity of $0.19_{-0.18}^{+0.05}$, and an $M \sin i = 0.46_{-0.13}^{+0.05} M_J$.}
\end{abstract}

\maketitle


\section{Introduction}

High-precision radial-velocity measurements of the
reflex radial velocity of host stars have lead to the discovery of over 150 planetary candidate
companions to solar-type stars both in single and multiple star systems (e.g., Eggenberger et al. 2004). Most of these planets have $M \sin i$ comparable to or greater than Jupiter, where $M$ is the mass of the planet and $i$ is the angle between the orbital plane and a plane perpendicular to the line of sight to the planetary system. However, as Doppler precision increases, lower mass planets are also being discovered and there are now five planets with an $M \sin i$ less than the mass of Neptune (Extrasolar Planets Encyclopaedia, http://vo.obspm.fr/exoplanetes/encyclo/index.php).

Progress is also being made in the development of improved algorithms for the analysis of these data. In Gregory (2005a) we described a Bayesian Markov chain Monte Carlo (MCMC) approach and demonstrated its effectiveness in the analysis of data from HD 73526 (Tinney et al. 2003). 
This approach has the following advantages:
\begin{itemize}
\item[a)] Good initial guesses of the parameter values are not required. A MCMC is capable of efficiently exploring all regions of joint prior parameter space having significant probability. There is thus no need to carry out a separate search for the orbital period(s).  
\item[b)] The method is also capable of fitting a portion of an orbital period, so search periods longer than the duration of the data are possible.
\item[c)] The built-in Occam's razor in the Bayesian analysis can save considerable time in deciding whether a detection is believable. More complicated models contain larger numbers of parameters and thus incur a larger Occam penalty, which is automatically incorporated in a Bayesian model selection analysis in a quantitative fashion. The analysis yields the relative probability of each of the models explored. 
\item[d)] The analysis yields the full marginal posterior probability density function (PDF) for each model parameter, not just the maximum {\it a posterior} (MAP) values and a Gaussian approximation of their uncertainties.
\item[e)] The inclusion of a novel statistical control system automates the otherwise time consuming process of selecting an efficient set of Gaussian parameter proposal distribution $\sigma$'s for use in the MCMC algorithm. 
\end{itemize}
In this paper we will describe a significant improvement to this algorithm and demonstrates its use in the re-analysis of data from HD 208487 obtained by Tinney et al. (2005). See Ford (2004) for an interesting application of MCMC methods to the interacting planetary system, GJ 876, in which mutual perturbations
are detectable that can be modeled with two precessing Keplerian orbits. 

\section{Algorithm Improvement}

Basically the algorithm was designed as a Bayesian tool for nonlinear model fitting. It is well known that nonlinear models can yield multiple solutions (peaks in probability) in the prior parameter space of interest. For the Kepler model with sparse data the target probability distribution can be very spiky. This is particularly a problem for the orbital period parameter, $P$, which spans the greatest number of decades. In general, the sharpness of the peak depends in part on how many periods fit within the duration of the data. The previous implementation of the parallel tempering algorithm employed a proposal distribution for $P$ which was a Gaussian in the logarithmic of $P$. This resulted in a constant fractional period resolution instead of a fixed absolute resolution, increasing the probability of detecting narrow spectral peaks at smaller values of $P$. However, this proved not to be entirely satisfactory because for the HD 73526 data set (Tinney et al. 2005) one of the three probability peaks (the highest) was not detected in two out of five trials (Gregory 2005a).

The latest implementation employs a proposal distribution in discrete frequency space with a frequency interval $\Delta \nu$ given by $\Delta \nu \times \mbox{data duration} = $ resolution (typically resolution $= 0.007$). The resolution corresponds to the fraction of the trial period that the last data point will be shifted by a change in search frequency of $\Delta \nu$. The width of any probability peak is directly related to this quantity so that a proposal distribution of this form corresponds more closely to achieving a uniform number of samples per peak. Using this search strategy on the HD 73526 data set all three peaks were detected in five out of five trials.

\section{Re-analysis of HD 208487}

With the exception of the change mentioned in the previous section, the details of the algorithm are as described in Gregory (2005a). Four different models ($M_{2j}, M_{2}, M_{1j}, \& \ M_{1s}$) were compared. $M_{1j} \& \ M_{2j}$ are one and two planet models which in addition to the known measurement errors include an empirical estimate of stellar jitter which is due in part to flows and inhomogeneities on the stellar surface (e.g., Wright 2005). For HD 208487 this is estimated to be somewhere in the range $2-4$ m s$^{-1}$. We used a value of 3 m s$^{-1}$. In model $M_{1s}$ we neglect the stellar noise jitter but incorporates an additional noise parameter, $s$, that can allow for any additional noise beyond the known measurement uncertainties~\negthinspace\footnote{In the absence of detailed knowledge of the sampling distribution for the extra noise, other than that it has a finite variance, the maximum entropy principle tells us that a Gaussian distribution would be the most conservative choice (i.e., maximally non committal about the information we don't have). We will assume the noise variance is finite and adopt a Gaussian distribution with a variance $s^2$. Thus, the combination of the known errors and extra noise has a Gaussian distribution with variance $= \sigma_i^2 + s^2$, where $\sigma_i$ is the standard deviation of the known noise for i$^{\mbox{\tiny th}}$ data point.}. For example, suppose that the star actually has two planets, and the model assumes only one is present. In regard to the single planet model, the velocity variations induced by the unknown second planet acts like an additional unknown noise term. Stellar jitter, if present, will also contribute to this extra noise term. In general, nature is more complicated than our model and known noise terms. Marginalizing $s$ has the desirable effect of treating anything in the data that can't be explained by the model and known measurement errors as noise, leading to the most conservative estimates of orbital parameters (see Sections 9.2.3 and 9.2.4 of Gregory (2005b) for a tutorial demonstration of this point). If there is no extra noise then the posterior probability distribution for $s$ will peak at $s = 0$. 

\begin{table}
\begin{tabular}{llll}
\hline
\tablehead{1}{l}{b}{Parameter\\}
  & \tablehead{1}{l}{b}{Prior\\}
  & \tablehead{1}{l}{b}{Lower bound\\}
  & \tablehead{1}{l}{b}{Upper bound\\}\\ 
\hline
$P_1 \ \& \ P_2$  (d) & Jeffreys & 0.5  & 4422  \\
$K_1 \ \& \ K_2$  (m s$^{-1}$) & Modified Jeffreys\footnote{Since the prior lower limits for $K$ and $s$ include zero, we used a modified Jeffrey's prior of the form
\begin{equation}
p(K|Model,I) = \frac{1}{K+K_a}\; \frac{1}{\ln\left(\frac{K_a + K_{\rm max}}{K_a}\right)}
\label{eq:orbit13}
\end{equation}
For $K \ll K_a$, $p(K|Model,I)$ behaves like a uniform prior and for $K \gg K_a$ it behaves like a Jeffreys prior. The $\ln\left(\frac{K_a + K_{\rm max}}{K_a}\right)$ term in the denominator ensures that the prior is normalized in the interval 0 to $K_{\rm max}$.} & 0 \ (K$_a = 1)$ & 400  \\
V  (m s$^{-1}$) & Uniform improper prior & all reals &   \\
$e_1 \ \& \ e_2$ & Uniform & 0 & 1 \\
$\chi_1 \ \& \ \chi_2$ & Uniform & 0 & 1 \\
$\omega_1 \ \& \ \omega_2$ & Uniform & 0 & $2 \pi$ \\
$s$  (m s$^{-1}$) & Modified Jeffreys & 0 \ (s$_a = 5$,\ typical measurement error) & 100  \\
\hline
\end{tabular}
\caption{Parameter prior probability distributions.}
\label{tab:priors}
\end{table}

The parameter prior probability distributions used in this analysis are given in Table~\ref{tab:priors}. For the $V$ parameter, we use a uniform prior which spans all real values and set $p(V|Model,I) = 1$. While this is an `improper' (non-normalized) prior, the normalization cancels when comparing different models, since it appears exactly once in each model. Note, that during the MCMC iterations there is nothing to constrain $P_1$ to be less than $P_2$. After the MCMC has completed the parameters are redefined so that $P_1 < P_2$.

Table~\ref{tab:Bayesfactors} shows the Bayes factors for the 4 models. The Bayes factor is just the ratio of the global likelihood of a particular model to model $M_{1j}$, taken as a reference. Section 5.3 of Gregory (2005a) describes how the Bayes factors are calculated. Assuming equal prior probability for the one and two planet models then Table~\ref{tab:Bayesfactors} indicates that a two planet model is favored over a one planet model. 
\begin{table}
\begin{tabular}{cccc}
\hline
\tablehead{1}{c}{b}{Model\\}
  & \tablehead{1}{c}{b}{Mean $s$\\($m\; s^{-1}$)}
  & \tablehead{1}{c}{b}{$p(D|{\rm Model},I)$\\}
  & \tablehead{1}{c}{b}{Bayes factor\\}\\ 
\hline
$M_{1j}$   &  &  $2.4 \pm 0.1 \times 10^{-55}$ & 1.0\\
$M_{1s}$   & 8 &  $7.0 \pm 0.4 \times 10^{-56}$ & 0.3\\
& & & \\
$M_{2}$  &  &  $6.5 \pm 0.7 \times 10^{-54}$ & 27.0\\
$M_{2j}$  &  &  $2.2 \pm 0.2 \times 10^{-54}$ & 9.1 \\
\hline
\end{tabular}
\caption{Bayes factors for model comparison.}
\label{tab:Bayesfactors}
\end{table}

\begin{table}
\begin{tabular}{lll|ll}
\hline
\tablehead{1}{l}{b}{Parameter\\}
  & \tablehead{1}{l}{b}{Tinney et al.\\(2005)}
  & \tablehead{1}{l}{b}{New\\analysis}
  & \tablehead{1}{l}{b}{Parameter\\}
  & \tablehead{1}{l}{b}{New\\analysis}\\
\hline
$P_1$  (d) & $130\pm1$ & $129.5_{-0.3}^{+0.8}$ & $P_2$  (d) & $998_{-62}^{+57}$ \\
& &  &  & \\
$K_1$  (m s$^{-1}$) & $20\pm2$ & $16.0_{-1.8}^{+1.5}$ & $K_2$  (m s$^{-1}$) & $9.8_{-2.7}^{+1.0}$  \\
& &  &  & \\
$e_1$ & $0.32\pm0.10$& $0.22_{-.07}^{+.14}$ & $e_2$ & $0.19_{-.18}^{+.05}$  \\
& &  &  & \\
$\omega_1$  (deg) & $126\pm40$& $126_{-39}^{+12}$ & $\omega_2$  (deg) & $157_{-60}^{+112}$ \\
& &  &  & \\
$a_1$  (AU) & $0.49\pm0.04$& $0.492_{-.001}^{+.002}$ & $a_2$  (AU) & $1.92_{-.08}^{+.07}$  \\
& &  &  & \\
$M_1 \sin i$  ($M_J$) & $0.45\pm0.05$& $0.37_{-.04}^{+.03}$ & $M_2 \sin i$  ($M_J$) & $0.46_{-.13}^{+.05}$ \\
& &  &  & \\
Periastron passage 1& $1002.8\pm10$& $1010_{-20}^{+8}$ & Periastron passage 2& $446_{-120}^{+186}$ \\
\ (JD - 2,450,000) & &  & \ (JD - 2,450,000) & \\
& &  &  & \\
RMS residuals (m s$^{-1}$) & 7.2 & 4.2 & & \\
& &  &  & \\
Reduced $\chi^2$  & 1.27 & 0.83 & & \\
\hline
\end{tabular}
\caption{Model parameter estimates}
\label{tab:parameters}
\end{table}

\begin{figure}[p]
  \includegraphics[height=0.9\textheight]{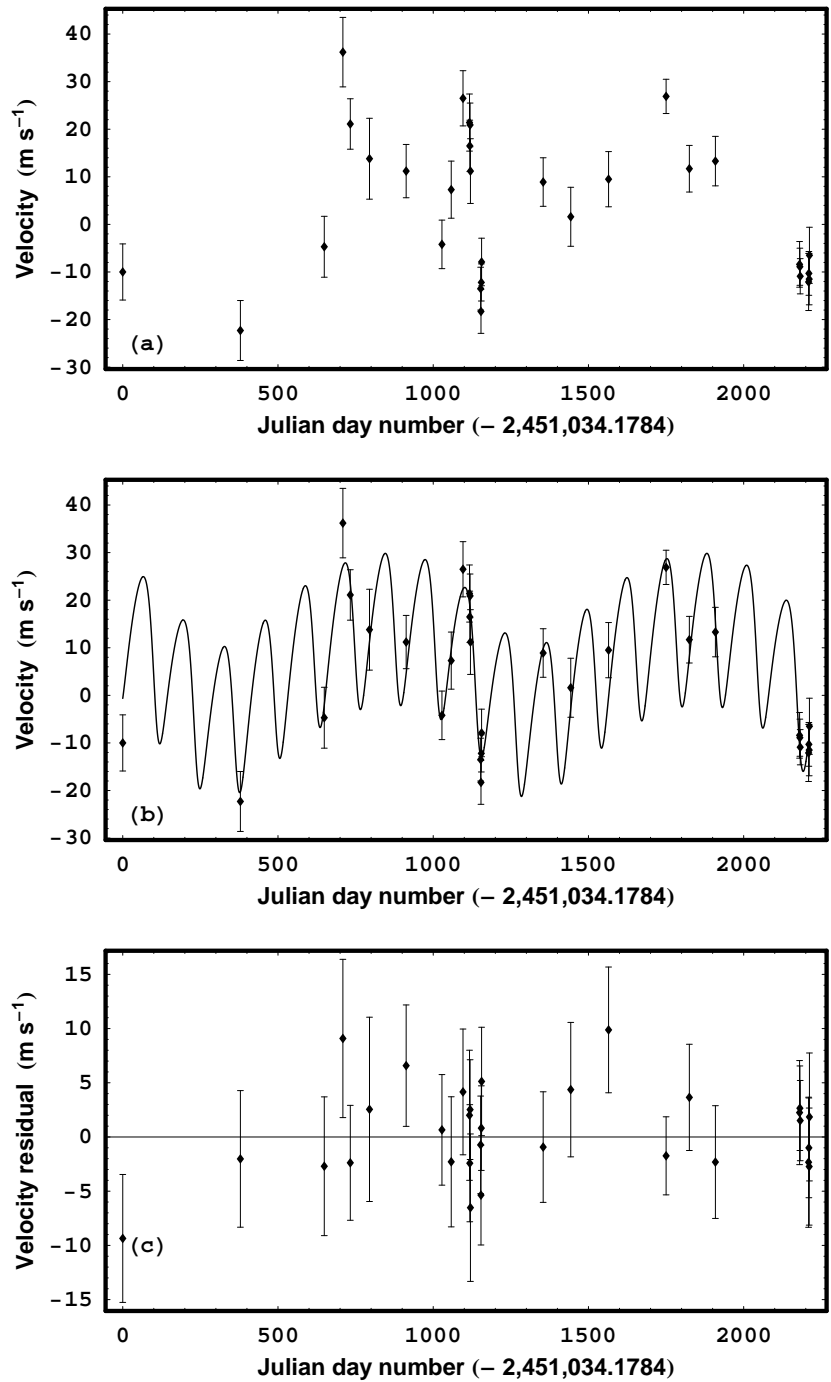}
  \caption{The raw data is shown in panel (a) and the best fitting two planet ($P_1 = 130$ day, $P_2 = 998$ day) model versus time is shown in (b). Panel (c) shows the residuals.}
\label{RawBestResid}
\end{figure}

\begin{figure}[h]
  \includegraphics[height=0.5\textheight]{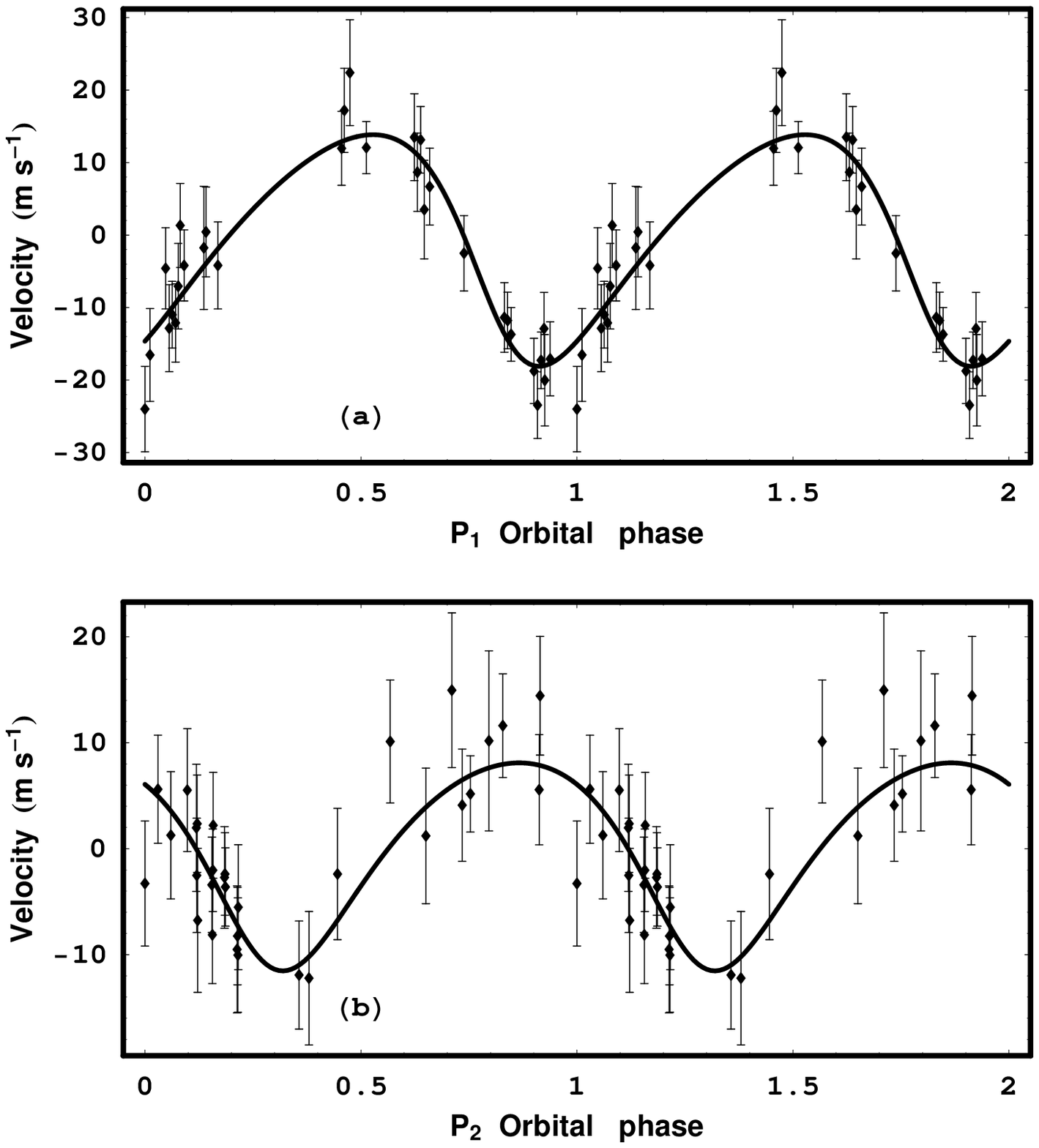}
  \caption{Panel (a) shows the data, with the best fitting $P_2$ orbit subtracted, for two cycles of $P_1$ phase with the best fitting $P_1$ orbit overlaid. Panel (b) shows the data plotted versus $P_2$ phase with the best fitting $P_1$ orbit removed.}
\label{P1P2Phase}
\end{figure}

The derived two planet model parameter values are given in Table~\ref{tab:parameters} and compared with the one planet parameters given in Tinney et al. (2005). Figures~\ref{RawBestResid} and \ref{P1P2Phase} show results for the most favored model $M_2$. The upper panel of Figure~\ref{RawBestResid} shows the raw data, the middle panel shows the best two planet fit and the lower panel shows the residuals which have an RMS $= 4.2$ m s$^{-1}$. Panel (a) of Figure~\ref{P1P2Phase} shows the data, minus the best fitting $P_2$ orbit, for two cycles of $P_1$ phase. The best fitting $P_1$ orbit is overlaid. Panel (b) shows the data plotted versus $P_2$ phase with the best fitting $P_1$ orbit removed.

\section{Conclusions}

In this paper we presented an improved version of the Bayesian parallel tempering MCMC algorithm of Gregory (2005a), which is a powerful Kepler periodogram for detecting multiple planets in precision radial velocity data. It can easily be modified to incorporate precessing Keperian orbits. It is also capable of fitting a portion of an orbital period, so early detection of periods longer than the duration of the data are possible. From a re-analysis of the HD 208487 data of Tinney et al (2005), we find strong evidence for a second planet with a period of $998_{-62}^{+57}$ days, an eccentricity of $0.19_{-0.18}^{+0.05}$, and an $M \sin i = 0.46_{0.13}^{0.05} M_J$.


\begin{theacknowledgments}
  This research was supported in part by grants from the Canadian Natural Sciences and Engineering Research Council at the University of British Columbia.
\end{theacknowledgments}





\end{document}